\documentclass[twocolumn]{aastex63}

\usepackage{amsmath}
\usepackage{amsbsy}

\newcommand\nh{\ifmmode{n_{\tiny \mbox{H}}}\else{$n_{\tiny \mbox{H}}$}\fi}
\newcommand\ngr{\ifmmode{n_{\tiny \mbox{gr}}}\else{$n_{\tiny \mbox{gr}}$}\fi}
\newcommand\jvsp{\ifmmode{j_{\tiny \nu,sp}}\else{$j_{\tiny \nu, sp}$}\fi}
\newcommand\jvff{\ifmmode{j_{\tiny \nu,ff}}\else{$j_{\tiny \nu,ff}$}\fi}
\newcommand\jvbb{\ifmmode{j_{\tiny \nu,bb}}\else{$j_{\tiny \nu,bd}$}\fi}
\newcommand\amax{\if{a_{\tiny \mbox{max}}}\else{$a_{\tiny \mbox{max}}$}\fi}
\newcommand\amin{\if{a_{\tiny \mbox{min}}}\else{$a_{\tiny \mbox{min}}$}\fi}
\newcommand\cmvol{\ifmmode{\mbox{cm}^{-3}}\else{$\mbox{cm}^{-3}$}\fi}
\newcommand\OIII{[$\textrm{O}\scriptstyle\mathrm{III}$]$_{88\mu m}$}
\newcommand\oiii{[$\textrm{O}\scriptstyle\mathrm{III}$]}
\received{June 1, 2019}
\revised{January 10, 2019}
\accepted{\today}
\submitjournal{ApJ}

\shorttitle{ALMA observation of the JWST high-$z$ galaxy candidate}
\shortauthors{Yoon}
\graphicspath{{./}{figures/}}

\begin{document}

\title{ALMA Observation of a $z\gtrsim10$ Galaxy Candidate Discovered with JWST}

\correspondingauthor{Ilsang Yoon}
\email{iyoon@nrao.edu}

\author[0000-0001-9163-0064]{Ilsang Yoon}
\affiliation{National Radio Astronomy Observatory, 520 Edgemont Road, Charlottesville, VA 22903, USA}

\author[0000-0001-6647-3861]{C.L. Carilli}
\affiliation{National Radio Astronomy Observatory, P.O. Box O, Socorro, NM, 87801, USA}

\author[0000-0001-7201-5066]{Seiji Fujimoto}\altaffiliation{Hubble Fellow}
\affiliation{Department of Astronomy, The University of Texas at Austin, Austin, TX, USA}
\affiliation{
Cosmic Dawn Center (DAWN), Jagtvej 128, DK2200 Copenhagen N, Denmark
}
\affiliation{
Niels Bohr Institute, University of Copenhagen, Lyngbyvej 2, DK2100 Copenhagen \O, Denmark
}

\author[0000-0001-9875-8263]{Marco Castellano}
\affiliation{INAF - Osservatorio Astronomico di Roma, via di Frascati 33,
00078 Monte Porzio Catone, Italy}

\author[0000-0001-6870-8900]{Emiliano Merlin}
\affiliation{INAF - Osservatorio Astronomico di Roma, via di Frascati 33,
00078 Monte Porzio Catone, Italy}

\author[0000-0002-9334-8705]{Paola Santini}
\affiliation{INAF - Osservatorio Astronomico di Roma, via di Frascati 33,
00078 Monte Porzio Catone, Italy}

\author[0000-0001-7095-7543]{Min S. Yun}
\affiliation{Department of Astronomy, University of Massachusetts, Amherst, MA 01003, USA}

\author[0000-0001-7089-7325]{Eric J.\,Murphy}
\affiliation{National Radio Astronomy Observatory, 520 Edgemont Road, Charlottesville, VA 22903, USA}

\author[0000-0003-1187-4240]{Intae Jung}
\affiliation{Space Telescope Science Institute, 3700 San Martin Drive, Baltimore, MD, 21218 USA}

\author[0000-0002-0930-6466]{Caitlin M. Casey}
\affiliation{Department of Astronomy, The University of Texas at Austin, Austin, TX, USA}

\author[0000-0001-8519-1130]{Steven L. Finkelstein}
\affiliation{Department of Astronomy, The University of Texas at Austin, Austin, TX, USA}

\author[0000-0001-7503-8482]{Casey Papovich}
\affiliation{Department of Physics and Astronomy, Texas A\&M University, College Station, Texas 77843, USA}

\author[0000-0003-3820-2823]{Adriano Fontana}
\affiliation{INAF - Osservatorio Astronomico di Roma, via di Frascati 33,
00078 Monte Porzio Catone, Italy}

\author[0000-0002-8460-0390]{Tommaso Treu}
\affiliation{Department of Physics and Astronomy, University of California, Los Angeles, 430 Portola Plaza, Los Angeles, CA 90095, USA}

\author{Jonathan Letai}
\affiliation{Department of Physics, Cornell University, 109 Clark Hall,
Ithaca, NY 14853, USA}

\begin{abstract}
We report the ALMA observation of a $z\gtrsim10$ galaxy candidate (GHZ1) discovered from the GLASS-JWST Early Release Science Program. Our ALMA program aims to detect the \oiii\ emission line at the rest-frame 3393.0062 GHz ($88.36\mu$m) and far-IR continuum emission with the spectral window setup seamlessly covering a 26.125 GHz frequency range ($10.10<z<11.14$). A total of 7 hours of on-source integration was employed, using four frequency settings to cover the full range (1.7 hours per setting), with $0\arcsec.7$ angular resolution. No line or continuum is clearly detected, with a 5$\sigma$ upper limit of the line emission of 0.93 mJy beam$^{-1}$ at 25 km s$^{-1}$ channel$^{-1}$ and of the continuum emission of 30$\mu$Jy beam$^{-1}$. We report marginal spectral (at 225 km s$^{-1}$ resolution) and continuum features ($4.1\sigma$ and $2.6\sigma$ peak signal-to-noise ratio, respectively), within $0\arcsec.17$ from the JWST position of GHZ1. This spectral feature implies $z=10.38$ and needs to be verified with further observations. Assuming that the best photometric redshift estimate ($z=10.60^{+0.52}_{-0.60}$) is correct, the broadband galaxy spectral energy distribution model for the $3\sigma$ upper limit of the continuum flux from GHZ1 suggests that GHZ1 has a small amount of dust ($M_d\lesssim10^4 M_{\odot}$) with high temperature ($T_d\gtrsim90$K). The $5\sigma$ upper limit of the \OIII\ line luminosity and the inferred star formation rate of GHZ1 is consistent with the properties of the low metallicity dwarf galaxies. We also report serendipitous clear detections of six continuum sources at the locations of the JWST galaxy counterparts in the field.
\end{abstract}



\section{Introduction} \label{sec:intro}
With the recent commissioning and the operation of JWST, the galaxy formation study enters into a new era: the JWST early release science (ERS) programs are discovering many galaxy candidates at $z\gtrsim10$ \citep[e.g.,][]{adams_etal_2022,atek_etal_2022,castellano_etal_2022,finkelstein_etal_2022,labbe_etal_2022,donnan_etal_2022,harikane_etal_2022b,naidu_etal_2022a,whitler_etal_2022,yan_etal_2022}. Although the majority of them is not spectroscopically confirmed yet, the rate of discovery has been prodigious and has opened up a new uncharted territory.

The unexpectedly large number of $z\gtrsim10$ galaxy candidates reported recently has already started to challenge current galaxy formation models \citep{boylan-kolchin_2022,ferrara_etal_2022,mason_etal_2022} and the cosmology \citep{lovell_etal_2022,menci_etal_2022}. The next key observational step is to confirm the spectroscopic redshift of those $z\gtrsim10$ galaxy candidates, to allay concerns that some $z\gtrsim10$ candidates may indeed turn out to be a low redshift ($z=4\sim5$) dusty star-forming galaxies \citep{naidu_etal_2022b,zavala_etal_2022} and bias our inference on the galaxy formation in the early Universe.  

The spectroscopic confirmation of the $z\gtrsim10$ galaxy candidates can be done using JWST NIRSpec by observing the rest-frame optical emission lines (with the degraded sensitivity increasingly significant in the long wavelength spectral configuration), or more easily by observing the rest-frame ultraviolet (UV) continuum break and emission lines, as shown by the recent observations of several $z\gtrsim10$ galaxies \citep{bunker_etal_2023,curtis-lake_etal_2022,roberts-borsani_etal_2022}.  

Another method to obtain the spectroscopic redshifts of those JWST-discovered $z\gtrsim10$ galaxies is to observe the fine-structure cooling lines at rest-frame far-IR (FIR) wavelengths. In particular, the \oiii\ emission line at the rest-frame 88.36$\mu$m (hereafter, \OIII) can be a good redshift marker for the high-$z$ galaxies \citep[e.g., $z=9.1$ galaxy from][]{hashimoto_etal_2018} and tends to be brighter with increasing redshift \citep{carniani_etal_2020,harikane_etal_2020}. Currently, there is no clear detection of the \OIII\ emission line from $z\gtrsim10$ galaxies except for the two recent claims of tentative emission: a $\sim 4\sigma$ signature from an $H$-dropout galaxy, HD1 \citep[$z=13.27$ from][]{harikane_etal_2022a} detected by Subaru Hyper-Suprime-Cam and a $\sim5\sigma$ signature found $0.\arcsec5$ away from the JWST detected galaxy, GHZ2 \citep{bakx_etal_2022}. If \OIII\ emission is observable from these JWST-discovered $z\gtrsim10$ galaxies, one can spectroscopically confirm the redshift and, if the FIR continuum is also detected, the galaxy spectral energy distribution (SED) model can be constrained much better with the inferred dust properties. 

We observed the rest-frame FIR continuum and \OIII\ emission from one of the recently discovered $z\gtrsim10$ galaxy candidates from the GLASS-JWST Early Release Science (ERS) program \citep{treu_etal_2022,castellano_etal_2022}, using Atacama Large Millimeter/submillimeter Array (ALMA). In this paper, we report the ALMA observation of our $z~\sim10.6$ target, GHZ1, and provide our analysis result on the ALMA data based on the standard LCDM cosmological model ($H_0=70$ km s$^{-1}$ Mpc$^{-1}$, $\Omega_{\mbox{\small m}}=0.3$, $\Omega_{\Lambda}=0.7$).

\section{Target: GHZ1} \label{sec:target}
The galaxy candidate targeted for our ALMA observation (2021.A.00023.S, PI: Yoon) was first discovered by two independent groups (designated as GHZ1 by \cite{castellano_etal_2022} and as GLz11 by \cite{naidu_etal_2022a}) from the JWST ERS program, GLASS-JWST \citep{treu_etal_2022} with the estimated photometric redshift (photo-$z$), $z\approx 10.6$ as suggested by strong Lyman break ($>2$ mag) feature at F115W and shorter. Multiple independent analyses confirm this galaxy as an ``ironclad'' candidate at $z\gtrsim10$ \citep{donnan_etal_2022,harikane_etal_2022b}. The target galaxy has an extended disk-like morphology with $10^{9.4\pm0.3} M_{\odot}$ stellar mass and 0.7 kpc half-light radius in the NIRCam F444W band image \citep{naidu_etal_2022a}. The improved calibration and updated photometry \citep{merlin_etal_2022} provide a slightly different stellar mass ($\mbox{log}_{10} M_{*}/M_{\odot}=9.1^{+0.3}_{-0.4}$ from \cite{santini_etal_2022}) and half-light radius ($0.5\pm0.02$ kpc from \cite{yang_etal_2022}),  If the redshift is confirmed spectroscopically, then this galaxy provides tantalizing evidence of galactic disk placed at $z\approx 11$ \citep{naidu_etal_2022a}. 

\section{ALMA observation}
\subsection{Design and Execution}\label{sec:design}
We design our observation to detect the rest-frame \OIII\ emission line and the FIR continuum emission. The reported photo-$z$ estimate by \cite{naidu_etal_2022a}: $z=10.62^{+0.27}_{-0.33}$ from EAZY \citep{brammer_etal_2008} and $z=10.86^{+0.52}_{-0.41}$ from Prospector \citep{leja_etal_2017}, has been revised based on the improved calibration of the NIRCam data and the updated photometry \citep{merlin_etal_2022,paris_etal_2023,weaver_etal_2023}, resulting in $z=10.60^{+0.18}_{-0.48}$ covering 50\% of the full photo-$z$ probability distribution \citep{merlin_etal_2022}. We set the center sky frequency (292.5 GHz) corresponding to  $z=10.60$ and construct four Science Goals (SG) with spectral setup such that the total 26.125GHz frequency range is covered without any frequency gaps (see Figure~\ref{fig:spectrum_full}). This results in the redshift coverage $10.10<z<11.14$ that contains 74\% of the photo-$z$ probability distribution. The observations were executed (two execution blocks for each SG) during 2022/09/17--2022/09/23 in excellent weather conditions (min/max PWV was 0.55/1.4 mm) with the C-3 configuration and 43--47 antennas. J2258-2758 was used for flux and bandpass calibration, and J2359-3133 was used for phase calibration. The resulting angular resolution with \texttt{robust}=0.5 is $0\arcsec.5$ (2 kpc at $z=10.6$) and the final angular resolution for the subsequent analysis with natural weighting (in Table~\ref{tab:obs}) is $0\arcsec.7$ (2.8 kpc at $z=10.6$) that is sufficiently large enough to enclose the spatially integrated flux from GHZ1 with 0.5 kpc half-light radius \citep{yang_etal_2022}.

\subsection{Data Calibration and Imaging}\label{sec:reduction}
We use \texttt{CASA} \citep{casa_2022} for data calibration and imaging. 
The standard ALMA pipeline (with \texttt{CASA} version 6.2.1-7) calibration has been applied. Continuum subtraction in $uv$-space was also done separately for each SG by the standard pipeline operation using a first-order baseline (i.e., linear baseline) and the pipeline identified continuum channels in the `dirty' cube. Then the calibrated visibility data is averaged in time (12.1 sec) and in spectral channel (5 channels) to create spectral line cube and continuum image. For the purpose of finding a faint emission line, we use the visibility data without continuum subtraction, so we avoid the noise rearrangement by continuum subtraction and preserve the original noise after the calibration. The subsequent imaging (for line and continuum) has been done with `auto-multithresh' algorithm \citep{kepley_etal_2020} for clean mask by the ALMA pipeline using natural weighting to increase the surface brightness sensitivity for detection. The final cube for analysis has 25 km s$^{-1}$ resolution. The achieved beam size and RMS sensitivity for the line emission with 25 km s$^{-1}$ resolution and for the aggregated continuum emission with the entire 26.125 GHz bandwidth, are summarized in Table~\ref{tab:obs}. Because of no strong signal observed in `dirty' cubes, the pipeline imaging did not perform a clean cycle for all cubes (as a result, all cubes in the subsequent analysis are `dirty' cubes without continuum subtraction). 

\subsection{Ancillary Spectral Cubes}\label{sec:othercubes}
We smooth our final 25 km s$^{-1}$ resolution cube using a boxcar smoothing kernel and create ancillary spectral cubes with 5 different kernel widths (3, 5, 7, 9, and 11 channels), using \texttt{CASA} task \texttt{specsmooth}. These ancillary spectral cubes are used to verify the robustness of the spectral feature that we find in the original cube (see Section~\ref{sec:tentative} for more detailed discussion). 

\begin{deluxetable}{ccccccc}
\tablecaption{ALMA Resolution and Sensitivity with Natural Weighting\label{tab:obs}}
\tablehead{\colhead{Beam (line)}           &
           \colhead{Beam (Cont.)}           &
           \colhead{\tablenotemark{$\dag$}RMS (line)}           &
           \colhead{RMS (Cont.)}     \\
           \colhead{FWHM}           &
           \colhead{FWHM}           &
           \colhead{$\mu$Jy/beam [$\Delta v$]}           &
           \colhead{$\mu$Jy/beam}
           }
\startdata
$0.84\arcsec\times0.65\arcsec$ & $0.80\arcsec\times0.60\arcsec$ & 186.0 [25 km s$^{-1}$] & 6.0 \\
\enddata
\tablenotetext{$\dag$}{Averaged over the frequential and spatial domain}
\end{deluxetable}

\begin{deluxetable*}{ccccccccccc}
\tablecaption{Properties of Line and Continuum Emission\label{tab:measure}}
\tablehead{\multicolumn{6}{c}{\oiii~(225 km s$^{-1}$ resolution)} &
           \multicolumn{4}{c}{Continuum} \\
           \colhead{Peak} &
           \colhead{RMS} &
           \colhead{\tablenotemark{$^\dag$}$S\Delta v$} &
           \colhead{\tablenotemark{$^\ddag$}$S\Delta v$} &
           \colhead{\tablenotemark{$^\dag$}$L_{[\mbox{\tiny OIII}]}$} &
           \colhead{\tablenotemark{$^\ddag$}$L_{[\mbox{\tiny OIII}]}$} &
           \colhead{Peak} &
           \colhead{RMS} &
           \colhead{\tablenotemark{$^\dag$}$F_{\mbox{\tiny 292.5GHz}}$} &
           \colhead{\tablenotemark{$^\ddag$}$F_{\mbox{\tiny 292.5GHz}}$} \\
           \colhead{mJy/beam} &
           \colhead{mJy/beam} &
           \colhead{Jy km s$^{-1}$} &
           \colhead{Jy km s$^{-1}$} &
           \colhead{L$_{\odot}$} &
           \colhead{L$_{\odot}$} &
           \colhead{$\mu$Jy/beam} &
           \colhead{$\mu$Jy/beam} &
           \colhead{$\mu$Jy} &
           \colhead{$\mu$Jy}
           }
\startdata
0.31  & 0.075  & 0.056  & $0.078\pm{0.023}$  & $2.06\times10^8$ & $2.83^{+0.83}_{-0.83}\times10^8$ & 15.6 & 6.0 & 18.0 & $6.3\pm{2.8}$\\
\enddata

\tablenotetext{$\dag$}{Based on $5\sigma$ upper limit for line (with 148 km s$^{-1}$ FWHM) and $3\sigma$ upper limit for continuum}
\tablenotetext{$\ddag$}{Based on the tentative line and continuum emission using, respectively, 0.\arcsec74 and 0.\arcsec35 radius circular aperture at the position of peak emission}
\end{deluxetable*}

\section{Result}\label{sec:result}
\subsection{\OIII\ emission}
\begin{figure*}
\includegraphics[width=1\textwidth]{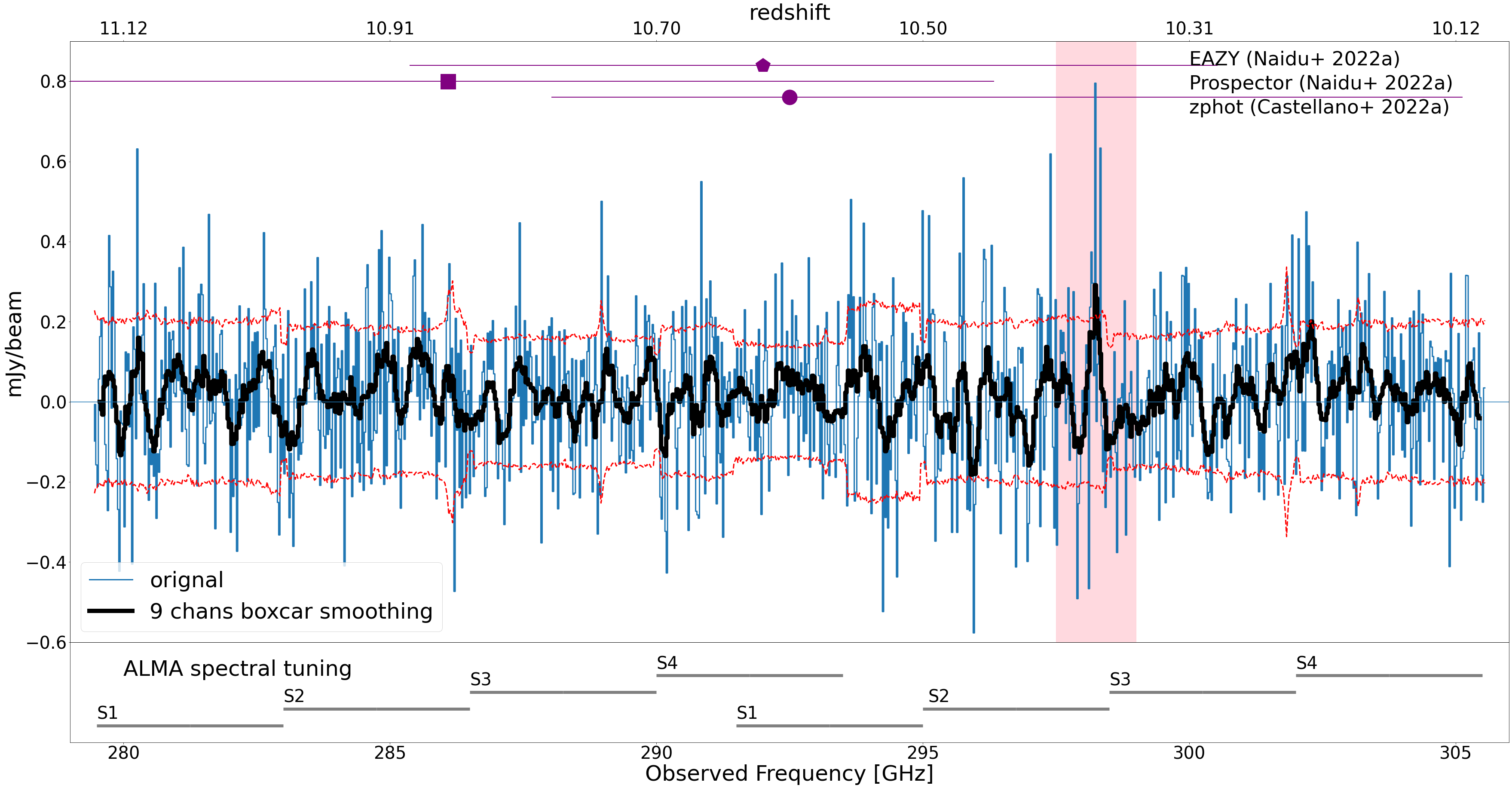}
\caption{ALMA spectrum with 25 km s$^{-1}$ resolution at the location of GHZ1 (extracted from a single voxel with a unit of mJy beam$^{-1}$) for the full 26.125 GHz spectral frequency range. Three photo-$z$ estimates are shown by the purple symbols and horizontal lines. The bottom panel shows the spectral setup of our ALMA observation seamlessly covering the entire frequency range. The spectrum is extracted from `dirty' cube without continuum subtraction. The blue line is the original spectrum and the black line is the spectrum with a boxcar smoothing using 9 channels. The red dashed line indicates $1\sigma$ RMS value at each spectral channel map and the spectrum highlighted by the pink shaded region that includes $\approx 4\sigma$ peak, is shown in Figure~\ref{fig:spectrum_zoomin}.}  
\label{fig:spectrum_full}
\end{figure*}

We extracted the spectrum at the location of GHZ1 from a single voxel in the spectral cube. Figure~\ref{fig:spectrum_full} shows the spectrum extracted from a single voxel at the location of GHZ1 (RA: 00:14:02.86 DEC: -30:22:18.70) for the entire 26.125 GHz frequency range (blue line for the original spectrum and black line for the spectrum with a boxcar smoothing using 9 channels). Three photo-$z$ estimates for GHZ1 are shown by the horizontal purple lines. The spectral setup (4 tunings with lower and upper sideband spectral windows) of our ALMA observation is shown in the bottom panel. The red dashed line indicates a $1\sigma$ RMS in each spectral channel map. We have investigated spectra at locations within one synthesized beam FWHM of the target galaxy and at spectral resolutions ranging from 75 km s$^{-1}$ channel$^{-1}$ to 275 km s$^{-1}$ channel$^{-1}$. No significant ($>5\sigma$) detection of \OIII\ emission is seen in the entire 26.125 GHz spectral cube over this region, with the $5\sigma$ upper limit at 25 km s$^{-1}$ channel$^{-1}$ being 0.93 mJy beam$^{-1}$. A marginal positive enhancement (later identified as a $4.1\sigma$ peak in the 225 km s$^{-1}$ resolution spectral channel map in Figure~\ref{fig:tentative}(a)) is seen in the spectrum at 298.25 GHz where there is no atmospheric absorption line and the spectral noise behaves smoothly across the frequency. 

\begin{figure}
\includegraphics[width=0.5\textwidth]{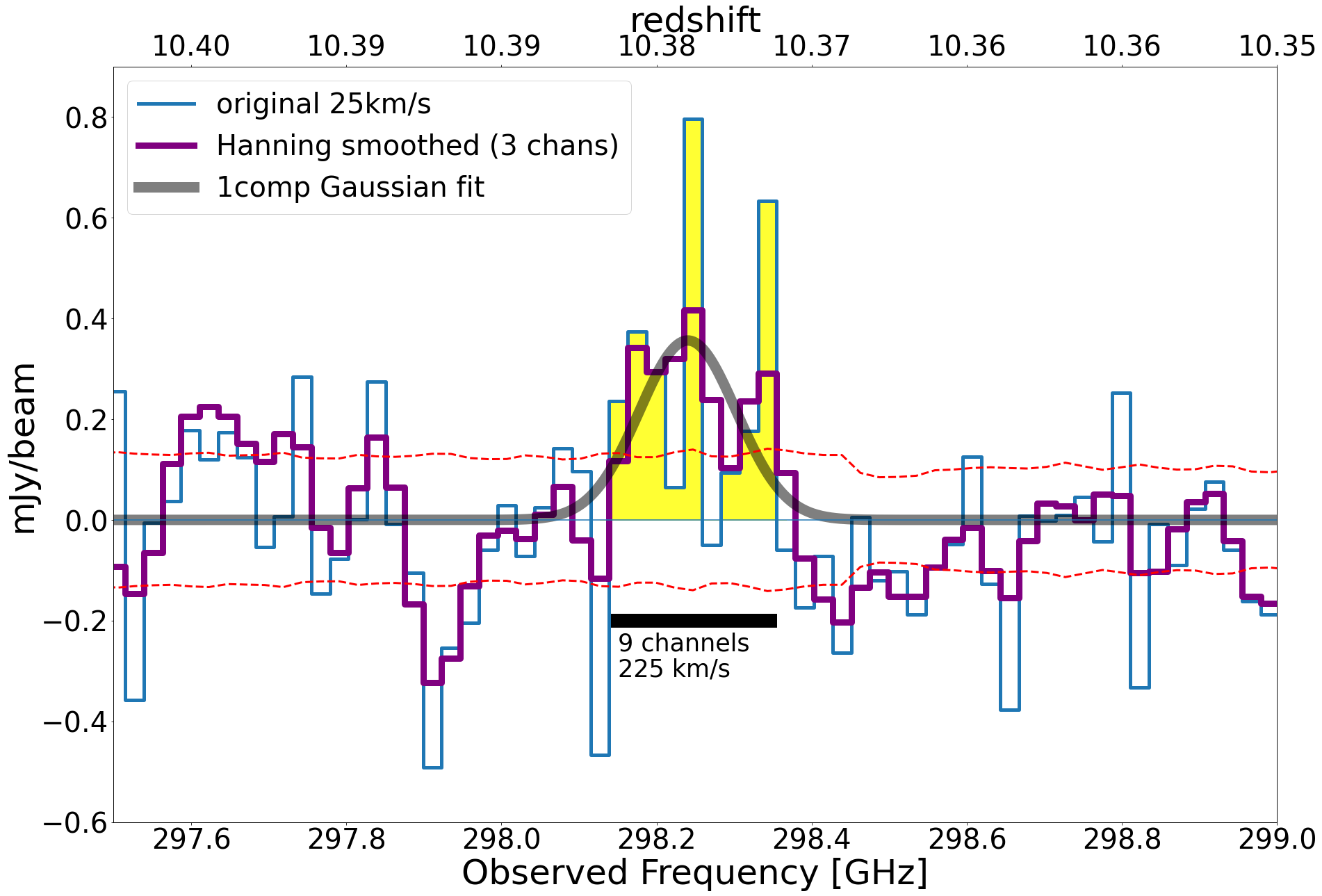}
\caption{The zoomed-in ALMA spectrum highlighted by pink-colored box in Figure~\ref{fig:spectrum_full}. In addition to the original spectrum (in blue), we show the Hanning smoothed spectrum (in purple) using 3 channels without decimating spectral channels. The red dashed line corresponds to $1\sigma$ RMS value at each spectral channel of the Hanning smoothed spectral cube. The black horizontal bar indicates the range of spectral channels (i.e., 9 channels) that is used to create a 225 km s$^{-1}$ resolution channel map in Figure~\ref{fig:tentative}(a). The thick solid gray line is the best fit 1-component Gaussian profile to the Hanning smoothed spectrum, with the velocity FWHM of 148 km s$^{-1}$.}
\label{fig:spectrum_zoomin}
\end{figure}

Figure~\ref{fig:spectrum_zoomin} shows a zoom-in view of the spectrum in the spectral region around this marginal enhancement (the pink shaded region in Figure~\ref{fig:spectrum_full}). In addition to the original 25 km s$^{-1}$ resolution spectrum (in blue), we also show the Hanning smoothed (using 3 channels) spectrum (in purple) with a $1\sigma$ RMS in each spectral channel map (red dashed line). If this feature is real, it would imply $z=10.38$. The black horizontal bar indicates the channels that were collapsed to create a 225 km s$^{-1}$ resolution channel map shown in Figure~\ref{fig:tentative}(a) and the gray thick solid line shows a fit to the Hanning smoothed spectrum with 1-component Gaussian profile. The peak of the Gaussian profile is at 298.24GHz ($z=10.38$) and its velocity FWHM is 148 km s$^{-1}$. We discuss this marginal spectral feature in Section~\ref{sec:tentative}.

\subsection{Thermal dust continuum emission}
The continuum emission at 292.5 GHz around the peak of the rest-frame FIR SED, is created by \texttt{CASA} multi-frequency synthesis imaging with \texttt{nterms}=1. We also do not find a significant ($>5\sigma$) continuum emission from GHZ1 while we observe a marginal $2.6\sigma$ continuum peak emission at the location of GHZ1 (Section~\ref{sec:tentative}). Interestingly, the FIR continuum is detected for other JWST galaxies in the ALMA field (Section~\ref{sec:othergal}). 

\subsection{Tentative line and continuum emission}\label{sec:tentative}
\begin{figure*}
\gridline{\fig{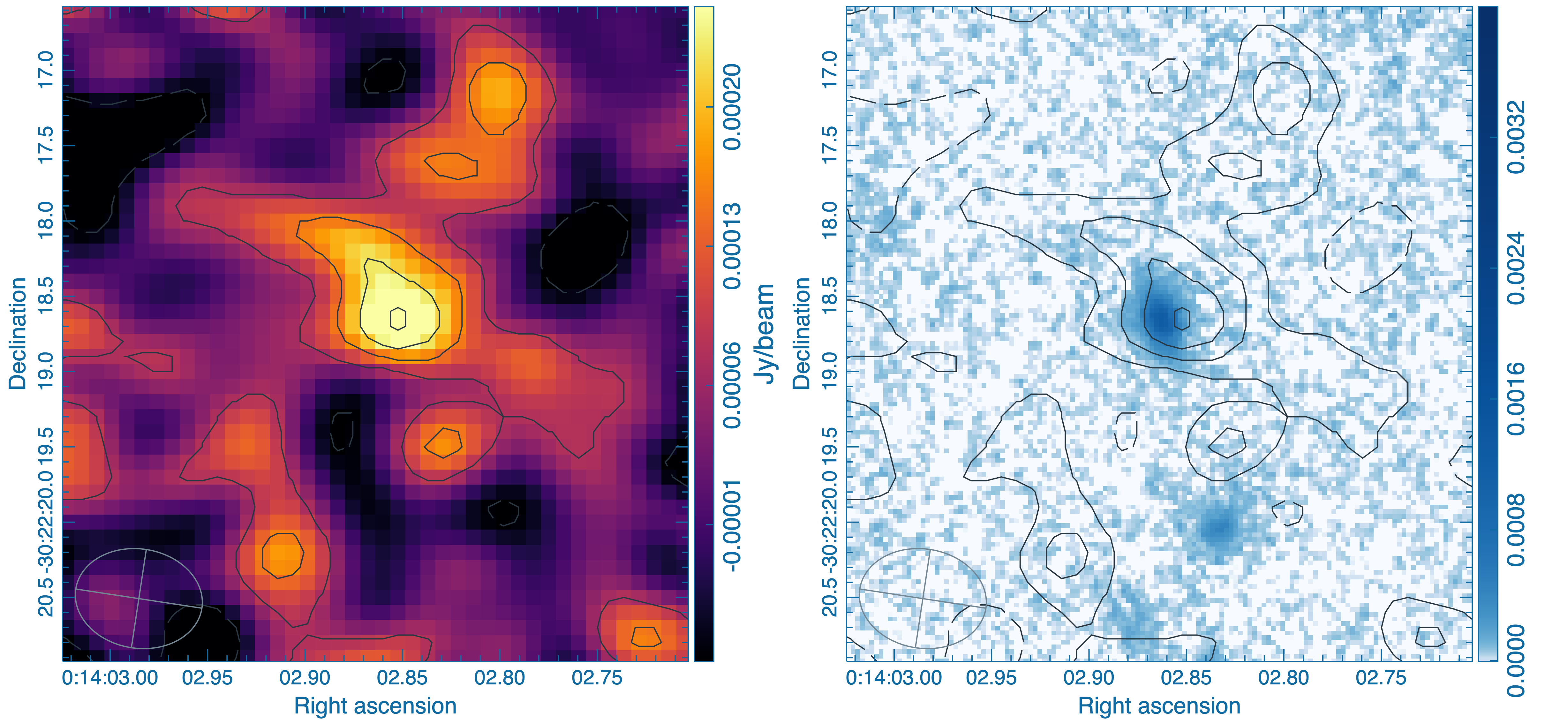}{0.54\textwidth}{(a) \OIII\ emission channel map}
\fig{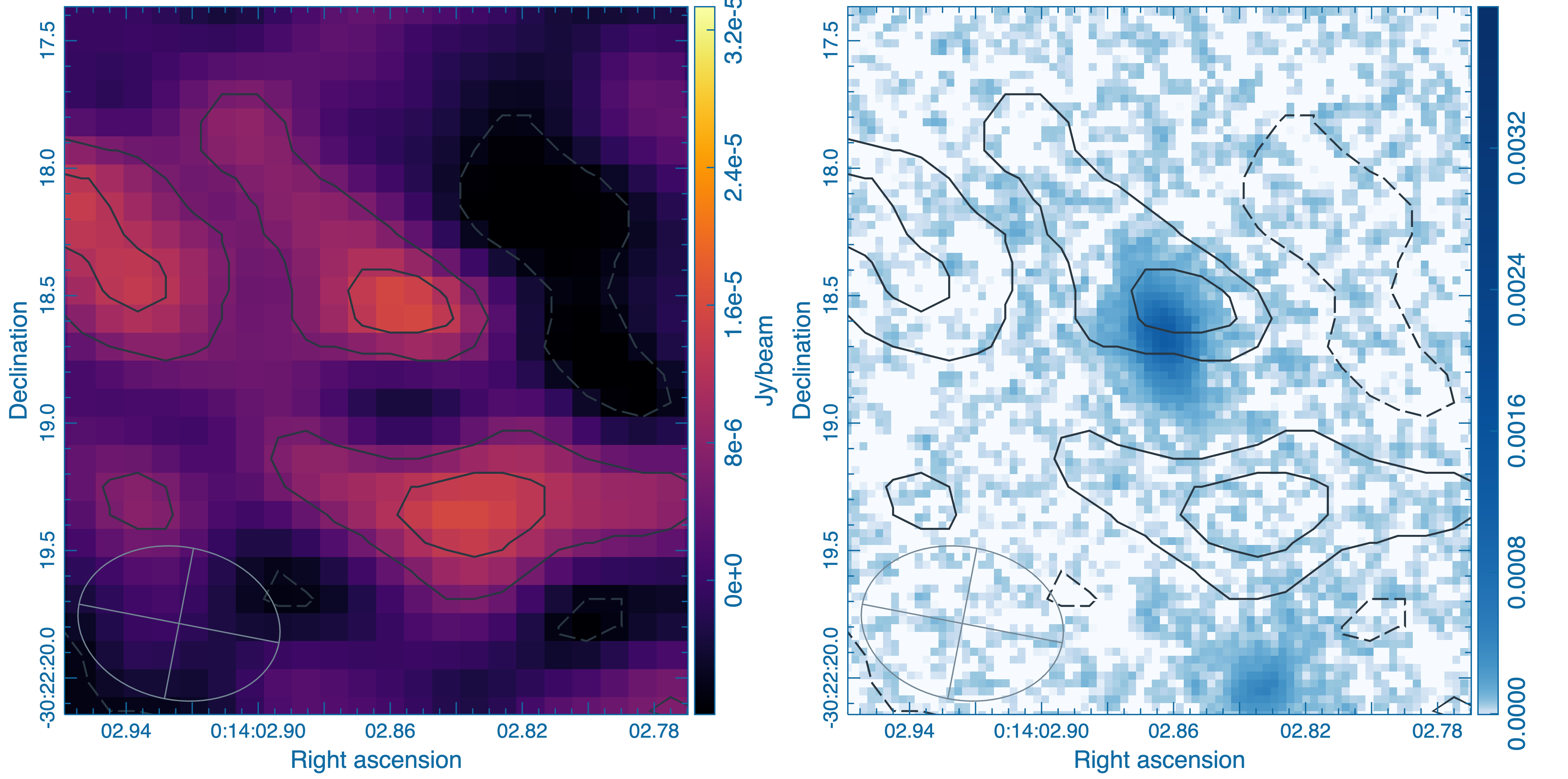}{0.5\textwidth}{(b) FIR continuum emission map}
}
\caption{Panel (a): 225 km s$^{-1}$ resolution channel map of the line feature associated with \OIII\ emission at 298.25 GHz observing frequency with -1 (dashed line), 1, 2, 3, 4 $\sigma$ contour (left) and JWST F444W NIRCam image of GHZ1 (right). The \textbf{$4.1\sigma$} peak emission (0.31 mJy/beam) is $\approx 1.5$ ALMA pixels ($0\arcsec.17$) away (to West) from GHZ1 shown in the JWST NIRCam image (right). Panel (b): FIR continuum map at 292.5 GHz observing frequency with -1 (dashed line), 1 and 2 $\sigma$ contours. The peak continuum emission ($15.6\mu$Jy/beam) is also $\approx 1.5$ ALMA pixels ($0\arcsec.17$) away (to North) from GHZ1 shown in the JWST NIRCam image (right). 
}
\label{fig:tentative}
\end{figure*}

Although we conclude that our ALMA observation does not detect a significant ($>5\sigma$) \OIII\ emission line and FIR continuum emission at the location of GHZ1, we find that there is a tentative `feature' in the spectral cube and continuum map that might be associated with GHZ1. In Figure~\ref{fig:tentative}(a), we show the 225 km s$^{-1}$ resolution channel map at 298.25 GHz using 9 channels shown in Figure~\ref{fig:spectrum_zoomin}. In Figure~\ref{fig:tentative}(b), we show our 26.125 GHz bandwidth continuum map created from all execution blocks in our observation. The measurements of the fluxes and RMS noises for the line and continuum map are summarized in Table~\ref{tab:measure}.

The collapsed channel map (Figure~\ref{fig:tentative}(a)) with $-1, 1, 2, 3, 4\sigma$ contour shows a $4.1\sigma$ peak emission (0.31 mJy beam$^{-1}$) at the location that is $0\arcsec.17$ away (to West) from GHZ1 seen in the JWST image on the right. The continuum map with $-1, 1, 2\sigma$ contour shows a $2.6\sigma$ peak emission (15.6 $\mu$Jy beam$^{-1}$) at the location that is also $0\arcsec.17$ away (to North) from GHZ1.

The $0\arcsec.17$ spatial offset from the GHZ1 is just a little larger than the `expected' $1\sigma$ astrometric uncertainty of JWST pointing ($0.\arcsec152$). The nominal RMS of ALMA absolute positional accuracy becomes poor if the peak signal-to-noise ratio of the target is low, and for our observation ($0.\arcsec7$ beam and peak S/N$\sim3$), it can be as large as $0.\arcsec26$ based on the approximate relationship between the positional accuracy, beam size and target peak S/N \citep{alma_thb_2019}. Also, we found that the astrometric difference between ALMA and JWST is small (less than the size of one ALMA pixel, i.e., $<0.\arcsec1$) based on the positional offset between the peak of the brightest ALMA continuum source in the field (shown in Figure~\ref{fig:othergal} discussed in Section~\ref{sec:othergal}) and its JWST counterpart. Therefore we do not expect a systematic positional offset between ALMA and JWST, and this observed $0\arcsec.17$ offset is likely to be a result of the combination of the random astrometric uncertainty of JWST and ALMA.

Moreover, the observation of high-$z$ galaxies suggests that the \OIII\ emission \citep{carniani_etal_2017} as well as the dust continuum emission \citep{inami_etal_2022} can be spatially offset from the galaxy optical emission. Therefore, just based on the positional offset alone, we cannot exclude the association of this tentative `feature' with the galaxy GHZ1. 

However, as emphasized in the recent study of the analysis of the noise in the ALMA data cube, it is tempting to trust a $3\sim4\sigma$ peak at the position where we expect to find it and we may be mistaken without testing how likely the peak is to be a noise fluctuation \citep{kaasinen_etal_2022}. To assess the significance of the tentative emission seen in the collapsed map, with a $0.\arcsec17$ spatial offset from the location of GHZ1, we did a simple investigation. First, we smoothed the 25 km s$^{-1}$ original spectral cube using a boxcar kernel with 5 different resolutions: 3, 5, 7, 9, and 11 channels (75, 125, 175, 225, and 275 km s$^{-1}$) as described in Section~\ref{sec:othercubes}, and searched for the strongest feature around the location of GHZ1 within a radius of 1 FWHM of the beam (0.\arcsec7), for each spectral cube. We find that the spectral feature that we identified at 298.25 GHz in the original 25 km s$^{-1}$ resolution spectrum (Figure~\ref{fig:spectrum_full}) still remains to be the strongest emission for all 5 resolutions. We confirm that a $4.1 \sigma$ emission line (0.31 mJy beam$^{-1}$) at 225 km s$^{-1}$ resolution (Figure~\ref{fig:tentative}(a)) is the strongest emission, and this positive feature is 50\% higher than any negative feature in the spectra.  

Second, we compute the probability that a $4.1\sigma$ emission seen at 225 km s$^{-1}$ resolution channel map could be due to Gaussian noise within a radius of 1 FWHM of the beam (0.\arcsec7) from the galaxy position, by counting the number of independent samples (i.e., number of independent channels $\times$ number of independent beams) in frequency and spatially within the area with 1 beam radius ($\pi r_{\mbox{\tiny FWHM}}^2$) across the entire 26.125 GHz spectral range. The final cube with the full 26.125 GHz frequency range used for the analysis has 120 channels with 225 km s$^{-1}$ (0.216 GHz) per channel resolution and the spatial area for search is within a radius of 1 beam FWHM resulting in 3.14 beams. Therefore the total number of independent samples is $\approx 380$. The Gaussian probability for a noise feature above $4\sigma$ is $3.2\times10^{-5}$. The product is then 0.012, which corresponds to the probability of seeing a $\ge 4\sigma$ noise feature over the number of independent samples, within the area of 1 beam FWHM radius. Therefore, the $4.1\sigma$ tentative emission we see from $0.\arcsec17$ ($<1$ beam FWHM) away from GHZ1 has a 1.2\% probability of being due to random Gaussian noise. 

More rigorous analysis \citep[e.g.,][]{kaasinen_etal_2022} of the noise can be done to increase the reliability of the claim of detection, although, in general, an ALMA noise field in a region void of any significant emission (the brightest continuum emission from the source in the southern edge of the field seen in Figure~\ref{fig:othergal} is only 87$\mu$Jy/beam, or $1.1\sigma$ at 225 km s$^{-1}$ channel$^{-1}$), is decidedly Gaussian in nature, dictated by the thermal noise of the system. Indeed, we performed an exercise of counting voxels with $>4\sigma$ or $<-4\sigma$ flux and computing their fraction out of the total number of voxels in the entire cube and found that the fractions ($3.2\times10^{-5}$ for $>4\sigma$ voxels and $4.5\times10^{-5}$ for $<-4\sigma$ voxels) are close to the Gaussian probability of $4\sigma$ ($3.2\times10^{-5}$). However, there is also an increase in the required search space inherent in a multivariate analysis which includes \textbf{a} resolution in space and frequency. So while the 1.2\% Gaussian noise probability at a specific spatial and spectral resolution seems low, we consider it optimistic, and it is certainly not low enough to claim a detection, but warrants further investigation. 

\subsection{Serendipitous detection of other galaxies}\label{sec:othergal}
\begin{figure*}
\gridline{\fig{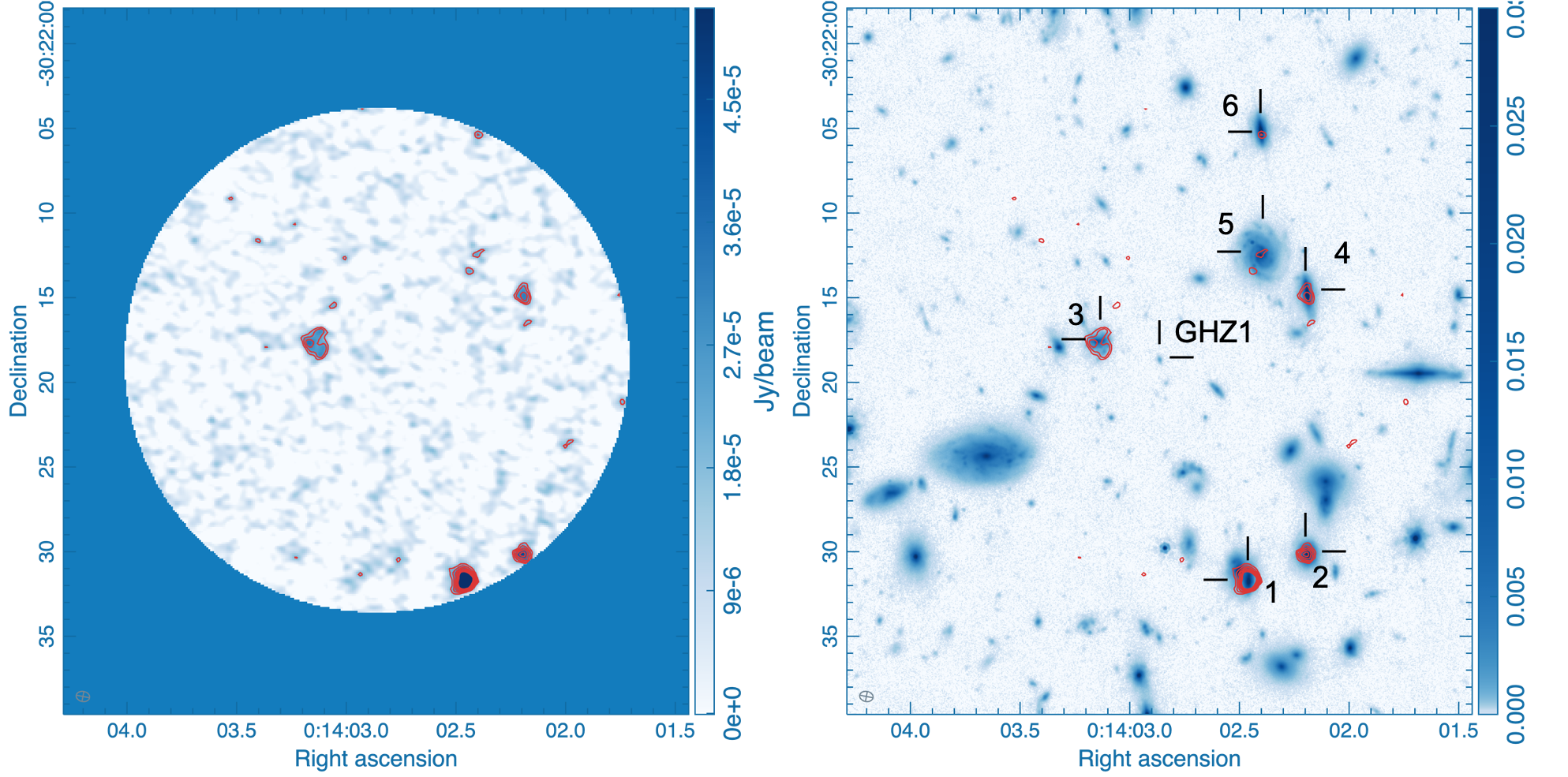}{1\textwidth}{}}
\caption{\textit{Left: }ALMA continuum image at 292.5 GHz with the overlayed red contours range from 3 to 8$\sigma$ level increased by $1\sigma$ \textit{Right: }JWST F444W NIRCam image of the galaxies in the ALMA field of view, with the contours of the ALMA continuum emission. ALMA detects FIR continuum emission from six galaxies denoted by numbers.}
\label{fig:othergal}
\end{figure*}

We create the continuum image by running \texttt{tclean} on the continuum visibility identified by the ALMA pipeline. We detected FIR continuum emission from six sources in the ALMA field of view and these six FIR sources have NIR counterparts in the JWST NIRCam image (Figure~\ref{fig:othergal}). The photometric redshift of these galaxies ranges from $z=0.6$ to $z=3$ \citep{merlin_etal_2022}. We found that 5 of them are not visible in the image created with \texttt{robust}=0.5, which indicates that the angular resolution for sufficient surface brightness sensitivity is very important to follow-up on the high-redshift JWST galaxies. The detailed information of the FIR flux and morphology of these sources, and the broadband SED analysis including the photometry from HST, JWST, and ALMA will be presented in a companion paper (Yoon et al. in preparation).

\section{Discussion}
We discuss the implications of the result from our observation. First, given the lack of significant ($>5\sigma$) \OIII\ emission line and thermal dust continuum emission from the location of GHZ1, we use the upper limit of the \OIII\ emission line and the continuum emission. Second, we use the measurement from the tentative emission. Lastly, we discuss the lessons from non-detection. 

\subsection{Upper-limit of [OIII] emission line}
We derive the \OIII\ luminosity from the velocity-integrated line flux density ($S \Delta v$ in Jy km s$^{-1}$) using $L_{\mbox{\scriptsize line}}=1.04\times10^{-3}S \Delta v D^2_{\mbox \tiny L} \nu_{\mbox{\scriptsize obs}} [L_{\odot}]$ \citep{carilli_walter_2013} where $D_{\mbox \tiny L}$ is luminosity distance in Mpc and $\nu_{\mbox{\scriptsize obs}}$ is observed frequency in GHz. For the line luminosity, we used the best photo-$z$ estimate ($z=10.60$ corresponding to 292.5 GHz observing frequency) and assume that the line is observed at the center frequency (292.5 GHz) of the full frequency range. The velocity integrated flux density ($S\Delta v$) and the \OIII\ luminosity ($L_{[\mbox{\tiny OIII}]}$) are in Table~\ref{tab:measure}.

Using the $5\sigma$ upper limit of \OIII\ luminosity ($2.06\times10^8 L_{\odot}$) with assumed 148 km s$^{-1}$ velocity FWHM from the Gaussian fit (Figure~\ref{fig:spectrum_zoomin}) and the star formation rate (SFR=$10.8^{+50.3}_{-6.1}$ M$_{\odot}$yr$^{-1}$) inferred from the galaxy SED modeling using NIRCam photometry \citep[][]{castellano_etal_2022b}, we present the SFR-$L_{[\mbox{\tiny OIII}]}$ relation in Figure~\ref{fig:sfr_o3}. The blue and orange line with a $1\sigma$ shaded-color region in Figure~\ref{fig:sfr_o3} shows the SFR-$L_{[\mbox{\tiny OIII}]}$ relation for low-metallicity dwarf galaxies and starburst galaxies \citep{delooze_etal_2014} and the red dot-dashed line is the best-fit relation for a handful of high-$z$ galaxies \citep{harikane_etal_2020}. The $5\sigma$ upper limit of GHZ1 is shown by the blue star symbol with the SFR error bar. The blue `x' symbol without an associated error bar shows the sum of `obscured' IR SFR \citep[using Equation (1) in ][]{hayward_etal_2014} determined by the IR luminosity output from the SED model and `unobscured' UV SFR determined by the UV luminosity \citep[using Equation (4) in][]{rosa-gonzlaez_etal_2002}. For comparison, we also plot the values of GHZ2 \citep{bakx_etal_2022}: $5\sigma$ upper limit at the location of GHZ2 in the JWST image (blue circle) and the $5.8\sigma$ detection from 0.\arcsec5 away from GHZ2 (red circle) with the SFR error bar. 

The \OIII\ upper limit for GHZ1 follows the SFR-$L_{[\mbox{\tiny OIII}]}$ relation for metal-poor dwarf galaxies. However, given that it is an upper limit, we cannot rule out a scenario that GHZ1 is consistent with the starburst galaxies (orange line in Figure~\ref{fig:sfr_o3}) although it is less likely.

\subsection{Upper-limit of thermal dust continuum emission}
Based on the $3\sigma$ upper limit of the 292.5 GHz continuum flux density and the JWST NIRCam photometry at F150W,F200W,F277W,F356W,F444W, we create the spectral energy distribution (SED) model of GHZ1 using \texttt{CIGALE} \citep{boquien_etal_2019} that computes a self-consistent SED from optical/NIR to FIR based on the amount of energy absorbed by dust (determined by dust extinction, $A_V=R_V E(B-V)$) and has been used to investigate the properties of dust in early galaxies \citep[e.g.,][]{burgarella_etal_2020}. We adopt the SED parameters ($e$-folding time scale for star formation history, stellar population age, stellar mass, $E(B-V)$) based on the NIRCam-only photometry \citep{merlin_etal_2022} for \texttt{CIGALE} and use single dust temperature graybody (dust emissivity $\beta=1.6$) model \citep{casey_2012} for FIR SED. 

For our modeling, we incorporate the systematic effects of CMB as a thermal background: an additional heating source and a reduced contrast against the background which become significant with increasing redshift as presented by the previous studies \citep[e.g.,][]{dacunha_etal_2013,Zhang_etal_2016}. Instead of adopting the conventional approach that uses FIR-only SED from a graybody function with a normalization from dust mass and compares it to the observed photometry corrected for the CMB effect, we choose to perform a full forward SED modeling. Although the full SED modeling based on energy balance has its own limitations due to parameter degeneracy and strong assumptions that may not always represent the reality \citep[e.g.,][]{buat_etal_2019}, it is still useful because the use of single FIR photometry cannot constrain the dust parameter (temperature and mass) and the JWST photometry helps to constrain the overall energy budget in FIR if the energy balance assumption (valid for compact and co-spatial IR and optical emission) holds for GHZ1, which is likely because of its extended regular disk-like morphology \citep{naidu_etal_2022a}.

Using \texttt{CIGALE}, we create a galaxy SED using graybody FIR spectrum with dust temperature increased by CMB heating following the same procedure in \cite{dacunha_etal_2013}. Since the `initial' model SED from \texttt{CIGALE} does not have the contribution from CMB heating, we add an additional graybody spectrum with the CMB temperature at a given redshift to the model SED and apply the correction factor due to the CMB contrast \citep{dacunha_etal_2013}. We note that the dust temperature and the mass are not independent, for the fixed IR luminosity (see Equation~\ref{eq:airtwo}): if we choose the dust temperature as a free parameter, the dust mass is determined, and vice versa. The detailed description of our SED modeling including the CMB effect is described in Appendix~\ref{sec:ap1}.

Figure~\ref{fig:sed} shows the SED models of GHZ1 obtained by varying SED with dust temperature and using $A_V=0.48$ that was chosen to match the observed JWST NIRCam photometry (black dots) by \cite{santini_etal_2022}. The total IR luminosity is determined by the dust-absorbed energy and the FIR SED shape is determined by the assumed dust temperature. Depending on the assumed dust temperature $T_d(0)$ without being affected by CMB heating at $z=10.60$, FIR SED can vary significantly without changing the NIR SED shape. Although it is not possible to obtain a reliable dust temperature by modeling a single FIR photometry, the best SED model prefers the FIR SED with a 90 K dust (with the corresponding $\approx 10^4 M_{\odot}$ dust mass) for our $3\sigma$ upper limit of FIR continuum emission at 292.5 GHz. 

\begin{figure}
\includegraphics[width=0.5\textwidth]{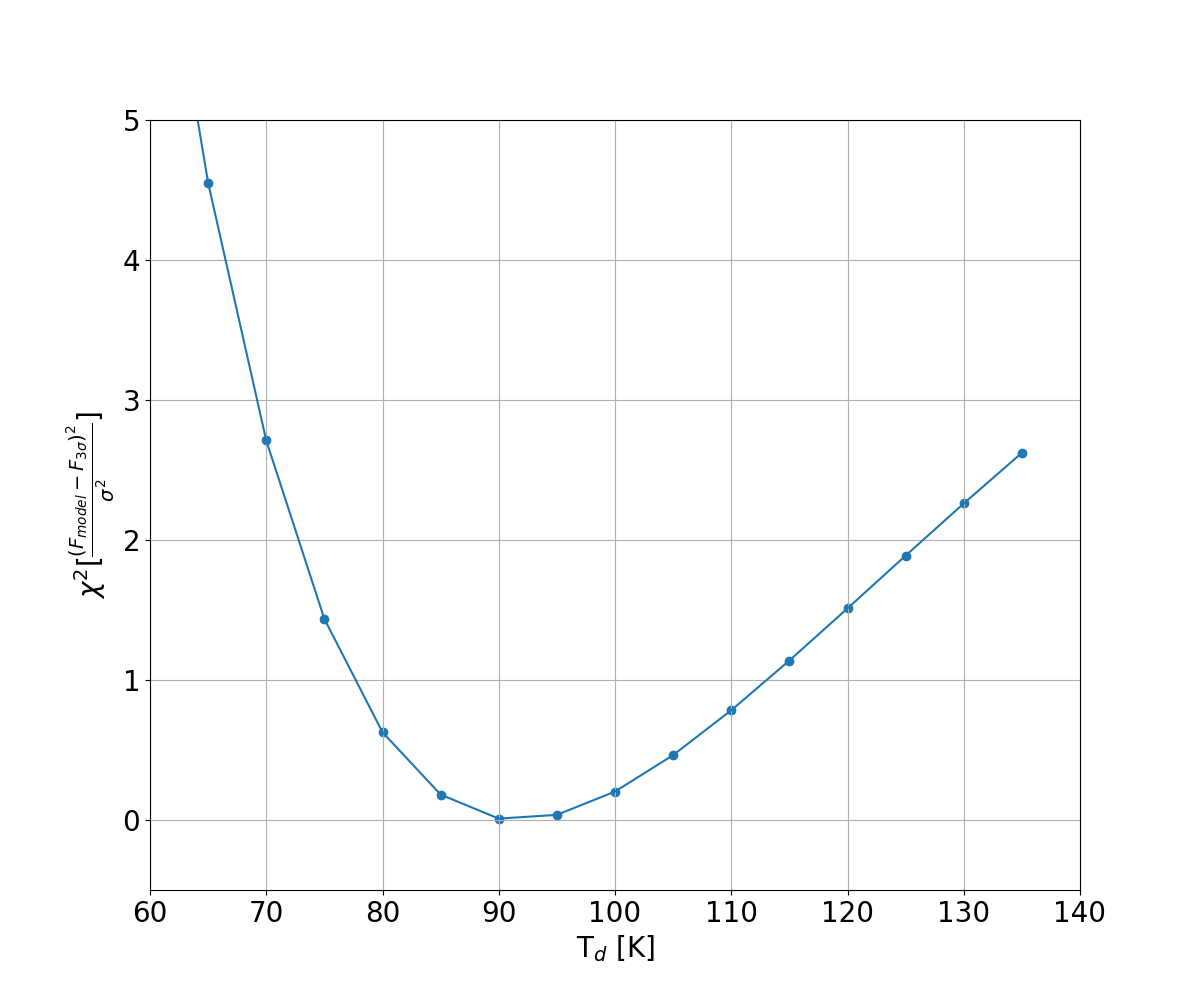}
\caption{$\chi^2$ value as a function of dust temperature. A single minimum value is observed at around $T_d(0)=90$K with a corresponding dust mass, $M_d=10^4 M_{\odot}$.}  
\label{fig:fom}
\end{figure}

In Figure~\ref{fig:fom}, we compute a $\chi^2$ value ($\chi^2=(F_{\mbox{\tiny model}}-F_{3\sigma})^2/\sigma^2$) of the model SED evaluated at the $3\sigma$ upper limit of the continuum flux density at 292.5 GHz, as a function of dust temperature. We note that the $\chi^2$ value estimated based on ``one'' data point does not have a strong statistical implication on the model parameter (i.e., dust temperature). However, Figure~\ref{fig:fom} shows that there is a single minimum value of $\chi^2$ at $T_d=90$K and the $3\sigma$ upper limit clearly disfavors the dust temperature that is significantly lower than 90K. 

The result implies that GHZ1 has very little dust with small ($\approx 10^{-5}$) dust-to-stellar mass ratio significantly lower than the typical values \citep[e.g., $10^{-2}\sim10^{-3}$ in][]{calura_etal_2017}. The small amount of dust in GHZ1 can be explained by dust ejection due to energetic star formation in the early Universe as suggested by the recent literature \citep[e.g.,][]{ferrara_etal_2022,fiore_etal_2022,nath_etal_2022,ziparo_etal_2022}. Also, the high dust temperature of GHZ1 is in line with the observational studies showing that the dust temperature of high-$z$ galaxies is higher than the local galaxies \citep[e.g.,][]{bakx_etal_2020,bakx_etal_2021,faisst_etal_2020,liang_etal_2019,schouws_etal_2022}, which supports for the low dust-to-gas ratio of high-$z$ galaxies \citep{hirashita_etal_2022}. Theoretical studies also suggest the increasing dust temperature with redshift \citep{behrens_etal_2018,sommovigo_etal_2021,sommovigo_etal_2022}. We note that the dust temperature estimate based on $3\sigma$ continuum upper limit is only a lower limit and can be even higher than 90K if the continuum emission is detected at a level below the current $3\sigma$ RMS.

\begin{figure}
\includegraphics[width=0.45\textwidth]{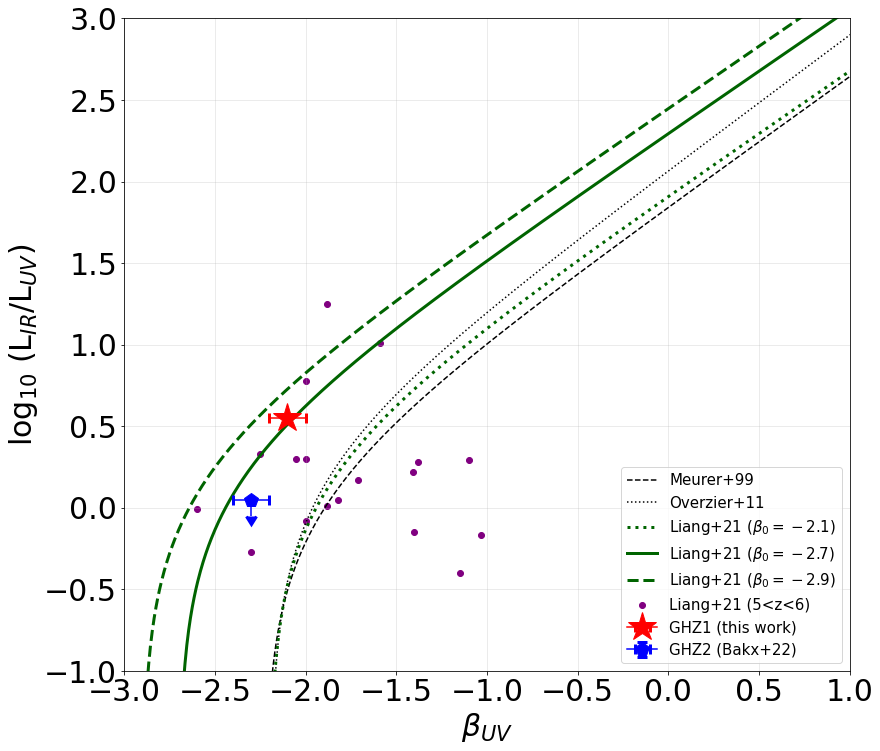}
\caption{IRX-$\beta$ relation for GHZ1 and GHZ2 with the distribution of the high-redshift ($5<z<6$) galaxies (purple dots) complied by \cite{liang_etal_2019} and their model relation with different intrinsic UV slopes ($\beta_0=-2.1, -2.7, -2.9$ shown by the green thick lines), and the reference relation from \cite{meurer_etal_1999} (black dashed line) and \cite{overzier_etal_2011} (black dotted line).}  
\label{fig:irxbeta}
\end{figure}

Also we locate GHZ1 in the IRX-$\beta_{\mbox{\tiny UV}}$ relation (Figure~\ref{fig:irxbeta}) where IRX$={L_{\mbox{\tiny IR}}}/{L_{\mbox{\tiny UV}}}$ and $\beta_{\mbox{\tiny UV}}$ is the UV continuum slope. The IR luminosity is estimated by integrating \texttt{CIGALE} SED for 8-1000$\mu$m rest-frame wavelength range and the UV luminosity is estimated based on $M_{\mbox{\tiny UV}}=-21.0$ \citep{naidu_etal_2022a,castellano_etal_2022}. We use $\beta_{\mbox{\tiny UV}}=-2.1\pm0.1$ \citep{naidu_etal_2022a}. Figure~\ref{fig:irxbeta} shows two IRX-$\beta_{\mbox{\tiny UV}}$ relations from \cite{meurer_etal_1999} (black dashed line) and \cite{overzier_etal_2011} (black dotted line) as a reference, and the distribution of high-redshift galaxies (purple dots) complied by \cite{liang_etal_2019}. The observed scatter of the galaxies in the IRX-$\beta_{\mbox{\tiny UV}}$ relation arises from the variations in the intrinsic UV spectral slope ($\beta_0$) and the shape (in particular, the slope \citep{salim_and_boquien_2019}) of the dust attenuation curve \citep[see ][and the reference therein]{salim_and_narayanan_2020}. A variation of the slope of dust attenuation curve changes the tilt of the IRX-$\beta_{\mbox{\tiny UV}}$ relation \citep[e.g.,][]{salim_and_boquien_2019} and a variation of the intrinsic UV slope shifts the IRX-$\beta_{\mbox{\tiny UV}}$ model curve horizontally \citep[e.g.,][]{liang_etal_2019}. For a simple dust slab model applied for simulated high-redshift galaxies, \cite{liang_etal_2019} found that the intrinsic UV slope ($\beta_0$) of galaxies largely explains the scatter in the IRX-$\beta_{\mbox{\tiny UV}}$ relation. 

In Figure~\ref{fig:irxbeta}, we also show three IRX-$\beta_{\mbox{\tiny UV}}$ relations from \cite{liang_etal_2019} using Milky Way dust attenuation curve with different $\beta_0$ (from the same value as the observed one to the bluer slope $\beta_0<-2.1$ as shown by green thick dotted, solid and dashed line). The locations of GHZ1 and GHZ2 in the IRX-$\beta_{\mbox{\tiny UV}}$ plane are shown by red star and blue pentagon symbol respectively. We find that the use of IRX-$\beta_{\mbox{\tiny UV}}$ relation in \cite{liang_etal_2019} with a bluer intrinsic UV slope (assumed to be $\beta_0=-2.7$) and the observed UV slope $\beta_{\mbox{\tiny UV}}=-2.1$ predicts the measured IRX value \footnote{The IR luminosity for the IRX estimate in this study is from the dust absorbed IR luminosity resulting from the best SED model, not from the arbitrary normalization by the assumed dust mass.} ($L_{\mbox{\tiny IR}}$ from \texttt{CIGALE} model SED and $L_{\mbox{\tiny UV}}$ from the observed JWST photometry) of GHZ1 well, which is consistent with the nature of GHZ1 (i.e., blue Lyman break galaxy at $z\approx10$).

\subsection{Measurement from the tentative line and continuum emission}
\begin{figure}
\includegraphics[width=0.5\textwidth]{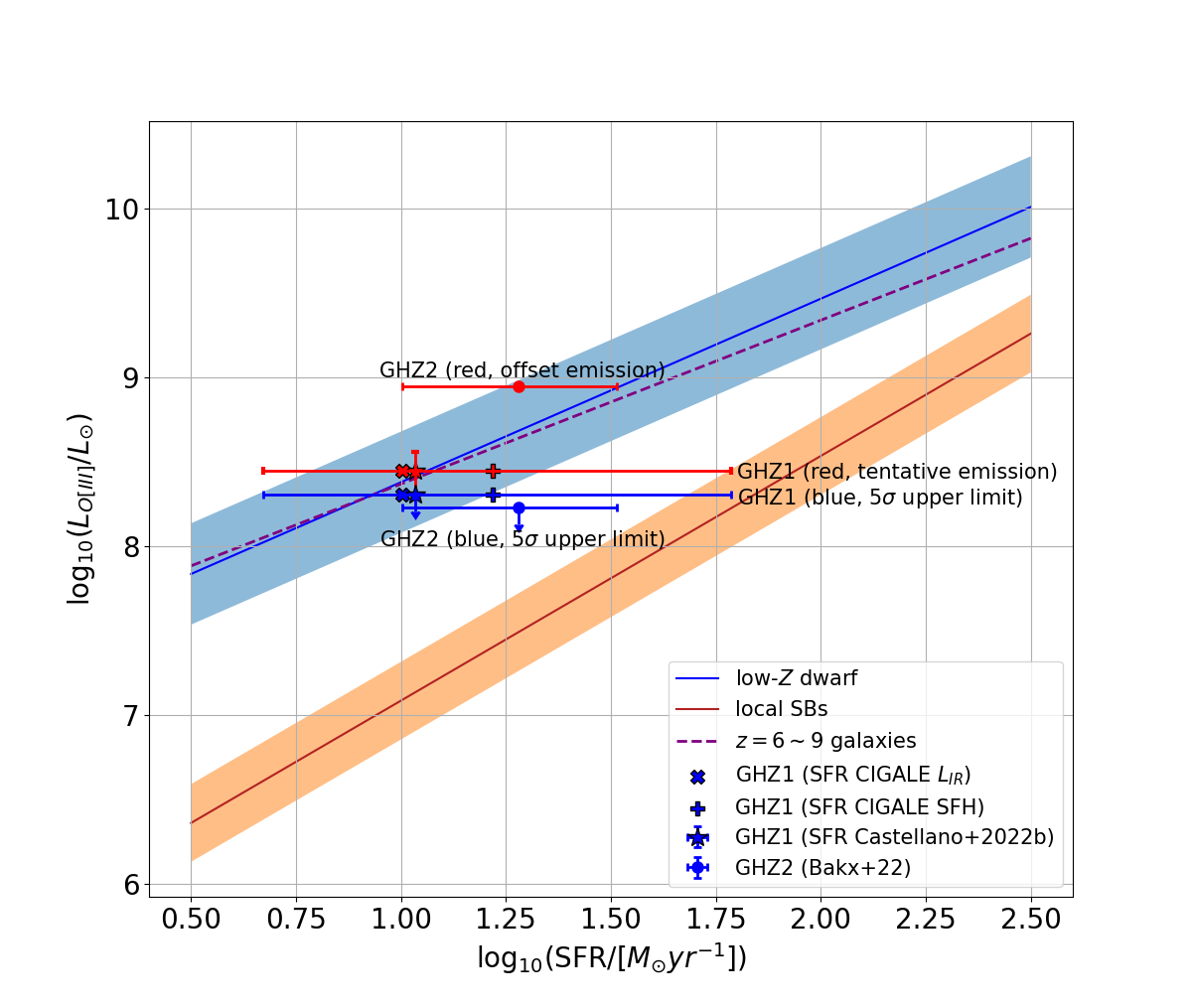}
\caption{The SFR and \OIII\ luminosity relation. The relation for the low metallicity dwarf galaxies (blue line) and for the starburst galaxies (orange line) with a $1\sigma$ shaded-color region are from \cite{delooze_etal_2014}, and the best-fit relation (purple dashed line) for a handful of high-redshift galaxies is from \cite{harikane_etal_2020}. Two galaxies observed by ALMA: GHZ2 by \cite{bakx_etal_2022} and GHZ1 (this work) with the $5\sigma$ upper limit of the \OIII\ luminosity, are shown by the black filled circle (GHZ2) and the black filled star (GHZ1). The \OIII\ luminosity for GHZ1 from the tentative emission is shown by the red filled star. The `x' symbol is for the SFR estimated by the sum of the SFR traced by IR (obscured SF) and UV (unobscured SF). The \OIII\ luminosity for GHZ2 from the offset position (0.\arcsec5 offset from the JWST position) is shown by the red filled circle.}  
\label{fig:sfr_o3}
\end{figure}

\begin{figure}
\includegraphics[width=0.5\textwidth]{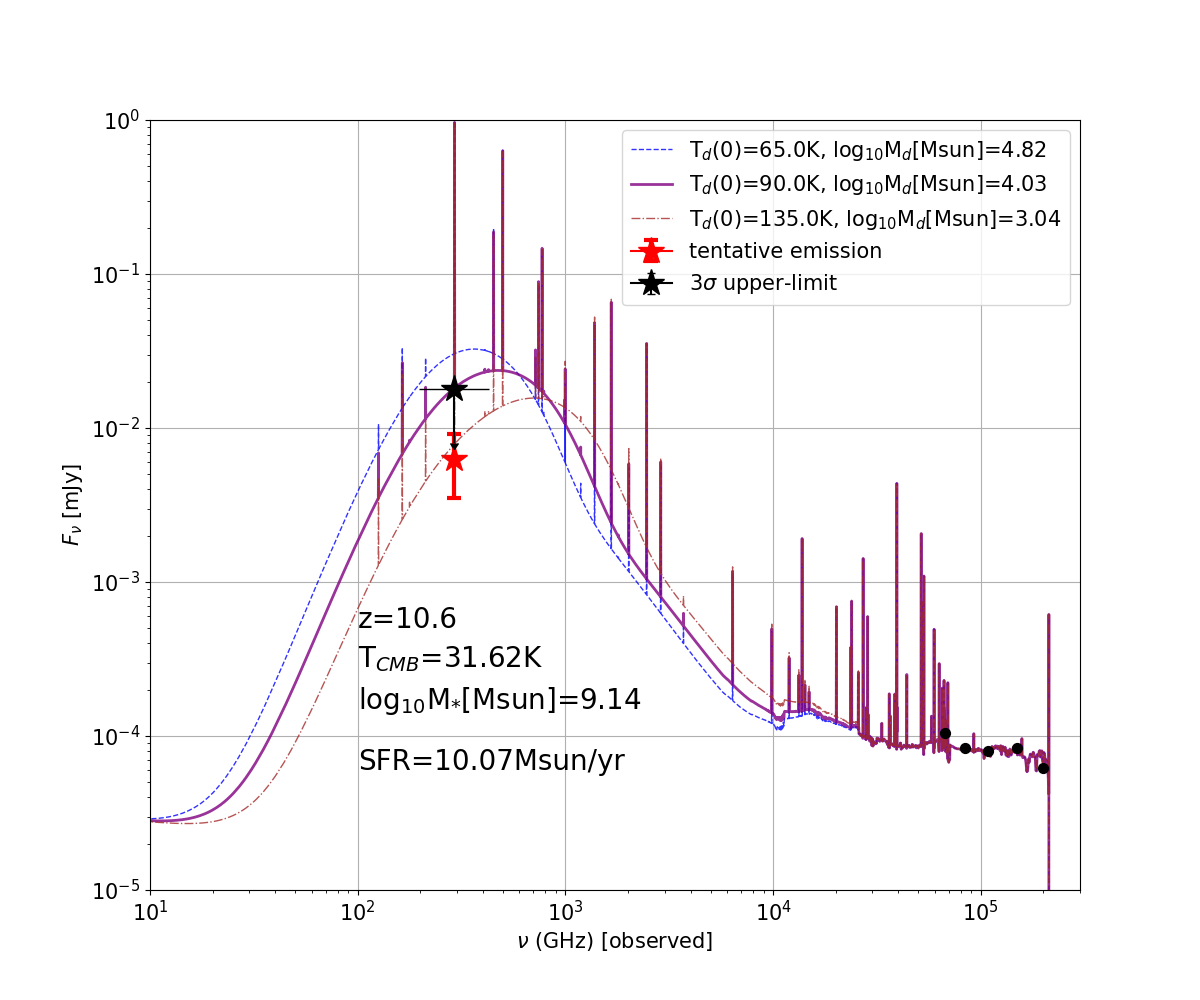}
\caption{SED models for GHZ1 based on the near-IR photometry (black dots) from JWST NIRCam and the $3\sigma$ upper limit (black star) of FIR photometry from ALMA, with the inclusion of CMB effect (see Appendix~\ref{sec:ap1} for more details). The best SED model suggests $T_d(0)=90$K and $M_d=10^4 M_{\odot}$.}  
\label{fig:sed}
\end{figure}

We also measure the line luminosity and the continuum flux density (reported in Table~\ref{tab:measure}) from the tentative feature in Section~\ref{sec:tentative}, using the line channel map and continuum map (Figure~\ref{fig:tentative}) and add them in Figure~\ref{fig:sfr_o3} and \ref{fig:sed}. The line measurement is performed using a circular aperture with a radius of $0.\arcsec74$ (comparable to $4\times$ALMA beam area) to enclose the extended emission and the continuum measurement is performed using a circular aperture with a radius of $0.\arcsec35$ (comparable to $1\times$ALMA beam area). The measurement error of the aperture photometry reported in Table~\ref{tab:measure} is the standard deviation of the flux density sampled with the same aperture from randomly selected 1000 positions (avoiding the region with continuum emission) in the field of view.

The \OIII\ luminosity of the tentative line emission at 298.25 GHz measured from the 225 km s$^{-1}$ resolution channel map (Figure~\ref{fig:tentative}(a)) is $2.83^{+0.83}_{-0.83}\times10^8 L_{\odot}$. The red star symbol with SFR and its error inferred from SED modeling \citep{castellano_etal_2022b} and the red `x' symbol with the sum of `obscured' and `unobscured' SFR in the SFR-$L_{[\mbox{\tiny OIII}]}$ relation (Figure~\ref{fig:sfr_o3}) are based on the \OIII\ luminosity from the tentative emission feature. They are consistent with the SFR-$L_{[\mbox{\tiny OIII}]}$ relation for the metal-poor dwarf galaxies, which may suggest that GHZ1 is metal-poor (also alluded by the low dust content inferred from the SED analysis) and has low ionization parameter that decreases the strength of \OIII\ emission \citep{kohandel_etal_2023}.

Modeling the Hanning smoothed spectrum (purple line in Figure~\ref{fig:spectrum_zoomin}) of the tentative line emission, using single Gaussian profile results in the center frequency of 298.24 GHz ($z=10.38$) and the velocity FWHM of 148 km s$^{-1}$ which is consistent with the FWHM values of the observed \oiii\ line (50-320 km s$^{-1}$) from other similarly high-redshift ($z\approx8-9$) galaxies \citep{inoue_etal_2016,laporte_etal_2017,laporte_etal_2021,hashimoto_etal_2018,tamura_etal_2019}. 

The continuum flux density from the tentative continuum emission (8.6 $\mu$Jy) is shown by the red star symbol (Figure~\ref{fig:sed}). The tentative continuum emission prefers even higher dust temperature ($T_d>90$K) or, in other words, smaller dust mass ($M_d<10^4 M_{\odot}$). The continuum upper limit and the tentative continuum emission suggest that given the non-negligible dust extinction value ($A_V=0.48$) to explain the JWST NIRCam photometry, the continuum flux density at 292.5GHz is lower than the expectation for a conventionally assumed dust temperature ($\approx 50$K) at high-$z$ Universe \citep[e.g., $z\approx7$ galaxies in][]{bakx_etal_2021,sommovigo_etal_2022}, which implies that the trend of increasing dust temperature at high-redshift may continue up-to $z\approx10$ and beyond.

\subsection{What we learn from a non-detection}
Recently another ALMA observation (2021.A.00020.S, PI: Bakx) of a $z\gtrsim10$ galaxy candidate, GHZ2, also reports the non-detection of \OIII\ emission and continuum emission \citep{bakx_etal_2022,popping_2022}. It turns out that the first two ALMA programs for spectroscopic confirmation of the redshift of these high-$z$ galaxy candidates discovered by JWST, do not detect a significant emission at the location of the galaxy. 

The non-detection of \OIII\ emission line can be explained by (1) insufficient line sensitivity, (2) `true' galaxy redshift much lower than $z\sim10$ (i.e., a low redshift interloper), or (3) the spectral coverage that is not wide enough to incorporate the full range of the photometric redshift probability distribution, as discussed by \cite{bakx_etal_2022} and \cite{popping_2022}. The second scenario can be ruled out for GHZ1 because the drop-out at F115W is too extreme, and the detected continuum is too blue, to be explained by the low-redshift Balmer break or strong dust obscuration of UV emission. The FIR continuum upper limit or the tentative continuum emission does not allow an extremely high dust extinction to mimic the F115W drop-out, using a low-$z$ SED model, which also helps to rule out the second scenario. The third scenario can be ruled out if ALMA has a wider-band and higher-sensitivity correlator in near future \citep{carpenter_etal_2020}. Like GHZ2, the insufficient sensitivity seems to be a plausible reason for the non-detection which can be explained by a low-metallicity interstellar medium \citep{bakx_etal_2022,popping_2022} or low-ionization parameter $U_{\mbox{\tiny ion}}$ in high-density environment \citep{kohandel_etal_2023}.

Like GHZ2, the non-detection of continuum emission from GHZ1 can be attributed to the low dust production rate \citep{bakx_etal_2022} or the dust ejection due to energetic feedback \citep{ferrara_etal_2022,fiore_etal_2022,nath_etal_2022,ziparo_etal_2022}. However, it is not clear what is responsible for the non-detection of the continuum emission from GHZ1. Given that the energy balance SED model with the upper limit possibly suggests a high temperature ($T_d>90$K) and small mass ($M_d<10^4 M_{\odot}$) dust, we need to have multiple FIR flux measurements at different frequencies to constrain the dust properties (mass, temperature, power-law slope of the Rayleigh-Jeans tail) of GHZ1 in order to understand the reason for the weak FIR continuum emission and more importantly to have an insight on the dust formation physics in the early Universe.

Based on our analysis, there is a tentative `feature' in the spectral cube around 298.25 GHz sky-frequency (suggesting $z=10.38$) and the continuum image, that is likely to be associated with GHZ1. Given the lack of sufficient sensitivity, a further follow-up observation combined with the current observation is necessary to increase the S/N and confirm the tentative `feature' in our data.

\section{Summary}\label{sec:summary}
Our ALMA DDT program (2021.A.00023.S) observed a $z\gtrsim10$ galaxy candidate, GHZ1, in the GLASS-JWST field to confirm its spectroscopic redshift using \OIII\ emission line. We find no clear $>5 \sigma$ detection of the line and continuum emission at the location of GHZ1, but report a tentative emission `feature' in the spectral cube ($4.1\sigma$ in the 225 km s$^{-1}$ resolution channel map) and continuum map ($2.6\sigma$) at a location close ($0.\arcsec17$ away) to GHZ1, that might be associated with GHZ1. If the line is real and identified at \OIII, it would imply $z=10.38$ which agrees with the photo-$z$ within its $1\sigma$ uncertainty. The \textbf{$5\sigma$} upper limit of the \OIII\ line emission and the inferred SFR suggest that GHZ1 in the SFR-$L_{[\mbox{\tiny OIII}]}$ plane is consistent with the metal-poor dwarf galaxies and the several observed high-$z$ galaxies. The SED modeling based on the \textbf{$3\sigma$} upper limit of the continuum emission and the JWST NIRCam photometry suggests that GHZ1 has a small fraction ($\lesssim10^{-5}$ of stellar mass) of hot dust ($T_d\gtrsim90$K). We need a confirmation of the tentative emission to have a firm conclusion. Finally, we also report six serendipitous FIR sources with the JWST galaxy counterpart, of which properties will be presented in a separate paper.

Although we report no clear detection of \OIII\ emission line, the initiative of the ALMA FIR observation of the $z\gtrsim10$ galaxy candidates discovered by JWST is an important step to improve our understanding of the galaxy population and ISM condition in the formation epoch of first galaxies. 

\software
{\texttt{CASA} \citep{casa_2022},
\texttt{carta} \citep{carta},
\texttt{numpy} \citep{harris_etal_2020}, 
\texttt{matplotlib} \citep{hunter_2007},
\texttt{CIGALE} \citep{boquien_etal_2019}}

\acknowledgments
We thank the anonymous referee for the constructive feedback that greatly improves the paper. I.Y. thank our ALMA P2G, Richard Simon for his help on the design of scheduling blocks of our ALMA observation. S.F. acknowledges support from the European Research Council (ERC) Consolidator Grant funding scheme (project ConTExt, grant No. 648179), the European Union’s Horizon 2020 research and innovation program under the Marie Sklodowska-Curie grant agreement No. 847523 ‘INTERACTIONS’, and the NASA Hubble Fellowship grant HST-HF2-51505.001-A awarded by the Space Telescope Science Institute, which is operated by the Association of Universities for Research in Astronomy, Incorporated, under NASA contract NAS5-26555. The Cosmic Dawn Center is funded by the Danish National Research Foundation under grant No. 140. This paper makes use of the following ALMA data: ADS/JAO.ALMA 2021.A.00023.S. ALMA is a partnership of ESO (representing its member states), NSF (USA) and NINS (Japan), together with NRC (Canada), MOST and ASIAA (Taiwan), and KASI (Republic of Korea), in cooperation with the Republic of Chile. The Joint ALMA Observatory is operated by ESO, AUI/NRAO and NAOJ. The National Radio Astronomy Observatory is a facility of the National Science Foundation operated under cooperative agreement by Associated Universities, Inc.

\appendix
\section{Computing Observed FIR SED with the inclusion of CMB effect}\label{sec:ap1}
The effect of CMB on the observation of thermal dust emission is twofold. CMB affects the FIR SED by enhancing the thermal blackbody component of the FIR SED (i.e., heating) but also by reducing the contrast of the detection (i.e., strong background emission). Therefore the `observed' FIR SED shape deviates from the $z=0$ SED with increasing redshift and the CMB effect (i.e., heating and contrast against the background) needs to be included when we model the SED. The basic theoretical framework and the prescription for the correction of CMB to the FIR SED model are described in \cite{dacunha_etal_2013}. Here we reformulate the prescription in \cite{dacunha_etal_2013} to include the normalization of IR luminosity and describe how we implement it into a panchromatic SED model (UV-to-FIR) from \texttt{CIGALE}.

To consider the CMB heating, we add an additional IR luminosity from dust heated by CMB at redshift $z$ to the modified blackbody SED component of FIR SED created by \texttt{CIGALE}. For high-redshift observations \citep{scoville_2013},
\begin{equation}
\nu_{obs}S_{\nu_{obs}} = \nu_{rest}\frac{L_{\nu_{rest}}}{4\pi D_L^2}
\end{equation} with $\nu_{rest}=\nu_{obs} \times (1+z)$. Using dust mass ($M_{dust}$) and mass absorption coefficient ($\kappa_\nu$), 
\begin{equation}
S_{\nu_{obs}}=(1+z)\kappa_{\nu_{rest}} 4\pi B_{\nu_{rest}}(T)\frac{M_{dust}}{4\pi D_L^2}
\end{equation}
Therefore,
\begin{equation}
L_{\mbox {\tiny IR}}=\int 4\pi D_L^2 S_{\nu_{obs}} d\nu_{obs} = \int \kappa_{\nu_{rest}} 4\pi B_{\nu_{rest}}(T)M_{dust} d\nu_{rest}
\end{equation}
where
\begin{equation}
\kappa_{\nu_{rest}} = \kappa_0 \left( \frac{\nu_{rest}}{\nu_0} \right)^{\beta} 
\end{equation} with the reference value $\kappa_0$ at a reference frequency $\nu_0$ that can adopted from observations. Then $L_{\mbox {\tiny IR}}$ from dust with temperature $T$ can be written as 
\begin{eqnarray}
L_{\mbox {\tiny IR}} & = & \int \kappa_{\nu_{rest}} 4\pi B_{\nu_{rest}}(T)M_{dust} d\nu_{rest} \nonumber \\
 & = & 4\pi \frac{2h\nu_0^4}{c^2} \left(\frac{kT}{h\nu_0}\right)^{4+\beta} \kappa_0 M_{dust} \int_{0}^{\infty} \frac{x^{3+\beta}}{e^x-1}dx
\end{eqnarray} where $x\equiv \frac{h\nu_{rest}}{kT}$. 
Riemann-zeta function $\zeta(s)$ is
\begin{equation}
\zeta(s) = \frac{1}{\Gamma(s)} \int_{0}^{\infty} \frac{x^{s-1}}{e^x-1}dx
\end{equation} where $\Gamma(s)=\int_{0}^{\infty} x^{s-1}e^{-x}$ is the gamma function. Therefore, using the values of the physical constants, the above equation is re-written as 
\begin{eqnarray}\label{eq:air}
L_{\mbox{\tiny IR}}  & = &  4\pi \frac{2h\nu_0^4}{c^2} \left(\frac{kT}{h\nu_0}\right)^{4+\beta} \kappa_0 M_{dust} \zeta(4+\beta)\Gamma(4+\beta) \nonumber \\
& = & 3.355\times10^{28}\left(\frac{\nu_0}{20.809~\mbox{\small GHz}}\right)^{-\beta}\left(\frac{T}{\mbox K}\right)^{4+\beta}\left(\frac{\kappa_0}{0.484~\mbox{cm}^2\mbox{g}^{-1}}\right) \left(\frac{M_{dust}}{M_{\odot}}\right)\zeta(4+\beta)\Gamma(4+\beta) \quad [\mbox {erg s}^{-1}]
\end{eqnarray} where $\nu_0$ is in GHz and $\kappa_0$ is in cm$^2$ g$^{-1}$. We adopt ${\nu_0}=352.697$GHz (or 850$\mu$m) and $\kappa_0=0.484$ cm$^2$ g$^{-1}$ from \cite{cochrane_etal_2022}. 
Using the emissivity index $\beta=1.6$, we obtain $L_{\mbox {\tiny IR}}$ in the unit of Watt (used in \texttt{CIGALE})
\begin{equation}\label{eq:airtwo}
L_{\mbox {\tiny IR}} = 2.29\times 10^{21}~\left(\frac{T}{\mbox K}\right)^{5.6} \left(\frac{M_{dust}}{M_{\odot}}\right) \quad [\mbox {W}]
\end{equation}
For the fixed $L_{\mbox {\tiny IR}}$, the higher the dust temperature is, the less the dust mass is.

If assuming thermal equilibrium, CMB provides additional heating to dust and this CMB heating becomes significant with increasing redshift. 
For CMB temperature at redshift, $T_{\mbox{\tiny CMB}} (z)$, the IR luminosity from CMB heating is 
\begin{equation}
L^{\mbox{\tiny CMB}}_{\mbox{\tiny IR}} = 2.29\times 10^{21}~\left(\frac{T_{\mbox{\tiny CMB}}(z)}{\mbox K}\right)^{5.6} \left(\frac{M_{dust}}{M_{\odot}}\right) \quad [\mbox {W}]
\end{equation} Because of the CMB heating, the dust temperature increases with redshift. Based \cite{dacunha_etal_2013}, we infer the dust temperature, $T_d (z)$ at a given redshift $z$ from
\begin{equation}\label{eq:at}
     T_d^{4+\beta}(z) = T_d^{4+\beta}(0)-T_{\mbox{\tiny CMB}}^{4+\beta}(0)+T_{\mbox{\tiny CMB}}(0)^{4+\beta}\times(1+z)^{4+\beta}
\end{equation} where $T_{\mbox{\tiny CMB}}(0)=2.725$K for the standard $\Lambda$CDM cosmology and $T_d(0)$ is the dust temperature at $z=0$.

For galaxy SED at $z$, we use \texttt{CIGALE} that assumes the energy balance: the energy created by the underlying stellar population is absorbed by the dust following the dust attenuation curve and the absorbed luminosity is re-emitted in FIR. For FIR SED module, we use a modified black body function with the elevated dust temperature $T_d(z)$ (Equation~\ref{eq:at}). However, we note that the resulting FIR SED from \texttt{CIGALE} does not include the CMB heating. We need to add $L^{\mbox{\tiny CMB}}_{\mbox{\tiny IR}}$ to the normalization of the \texttt{CIGALE} FIR SED which we can compute numerically. If we write the normalization of the \texttt{CIGALE} FIR SED for $T_d(z)$ without CMB heating included, $L^{\mbox{\tiny abs}}_{\mbox{\tiny IR}}$, the `boosted' IR SED can be obtained by correcting the \texttt{CIGALE} FIR SED with $T_d(z)$ by multiplying $(L^{\mbox{\tiny abs}}_{\mbox{\tiny IR}} + L^{\mbox{\tiny CMB}}_{\mbox{\tiny IR}})/L^{\mbox{\tiny abs}}_{\mbox{\tiny IR}}$.

Now the `boosted' FIR luminosity, $L^{\mbox{\tiny tot}}_{\mbox{\tiny IR}}$, (due to the CMB heating) with elevated dust temperature $T_d(z)$ is the sum of $L^{\mbox{\tiny abs}}_{\mbox{\tiny IR}}$ and $L^{\mbox{\tiny CMB}}_{\mbox{\tiny IR}}$ and can be written as 
\begin{eqnarray}
    L^{\mbox{\tiny tot}}_{\mbox{\tiny IR}} & = & 2.29\times 10^{21}~\left(\frac{T_d(z)}{\mbox K}\right)^{5.6} \left(\frac{M_{dust}}{M_{\odot}}\right) \nonumber \\
    & = & 2.29\times 10^{21}~\left(\frac{T_{\mbox{\tiny CMB}}(z)}{\mbox K}\right)^{5.6} \left(\frac{M_{dust}}{M_{\odot}}\right) + L^{\mbox{\tiny abs}}_{\mbox{\tiny IR}}
\end{eqnarray}
This results in a dust mass consistent with the luminosity balance: $L^{\mbox{\tiny tot}}_{\mbox{\tiny IR}}=L^{\mbox{\tiny CMB}}_{\mbox{\tiny IR}} + L^{\mbox{\tiny abs}}_{\mbox{\tiny IR}}$
\begin{equation}
    \left(\frac{M_{dust}}{M_{\odot}}\right) = \frac{L^{\mbox{\tiny abs}}_{\mbox{\tiny IR}}}{2.29\times 10^{21} \left[\left(\frac{T_d(z)}{\mbox K}\right)^{5.6} - \left(\frac{T_{\mbox{\tiny CMB}}(z)}{\mbox K}\right)^{5.6}\right]} 
\end{equation}

In our procedure for computing the `boosted' FIR SED by CMB heating, we numerically compute $L^{\mbox{\tiny abs}}_{\mbox{\tiny IR}}$ from the \texttt{CIGALE} FIR SED and estimate $M_{dust}$ for given $T_d(z)$ and $T_{\mbox{\tiny CMB}}(z)$. Then $L^{\mbox{\tiny CMB}}_{\mbox{\tiny IR}}$ is computed and the \texttt{CIGALE} FIR SED is multiplied by $(L^{\mbox{\tiny abs}}_{\mbox{\tiny IR}} + L^{\mbox{\tiny CMB}}_{\mbox{\tiny IR}})/L^{\mbox{\tiny abs}}_{\mbox{\tiny IR}}$. This is equivalent to scale the FIR SED model by multiplying a scale factor $\left(\frac{T_d(z)}{T_d(0)}\right)^{4+\beta}$ \citep{dacunha_etal_2013}.

The other aspect of CMB effect is the suppression of SED because the `observed' SED is against the background emission from CMB. Following \cite{dacunha_etal_2013}, we apply the correction factor, $1-\{B_\nu\left[T_{\mbox{\tiny CMB}}(z)\right]/B_\nu\left[T_{d}(z)\right]\}$ to the modified blackbody component of the `boosted' FIR SED.

\begin{figure}
\plotone{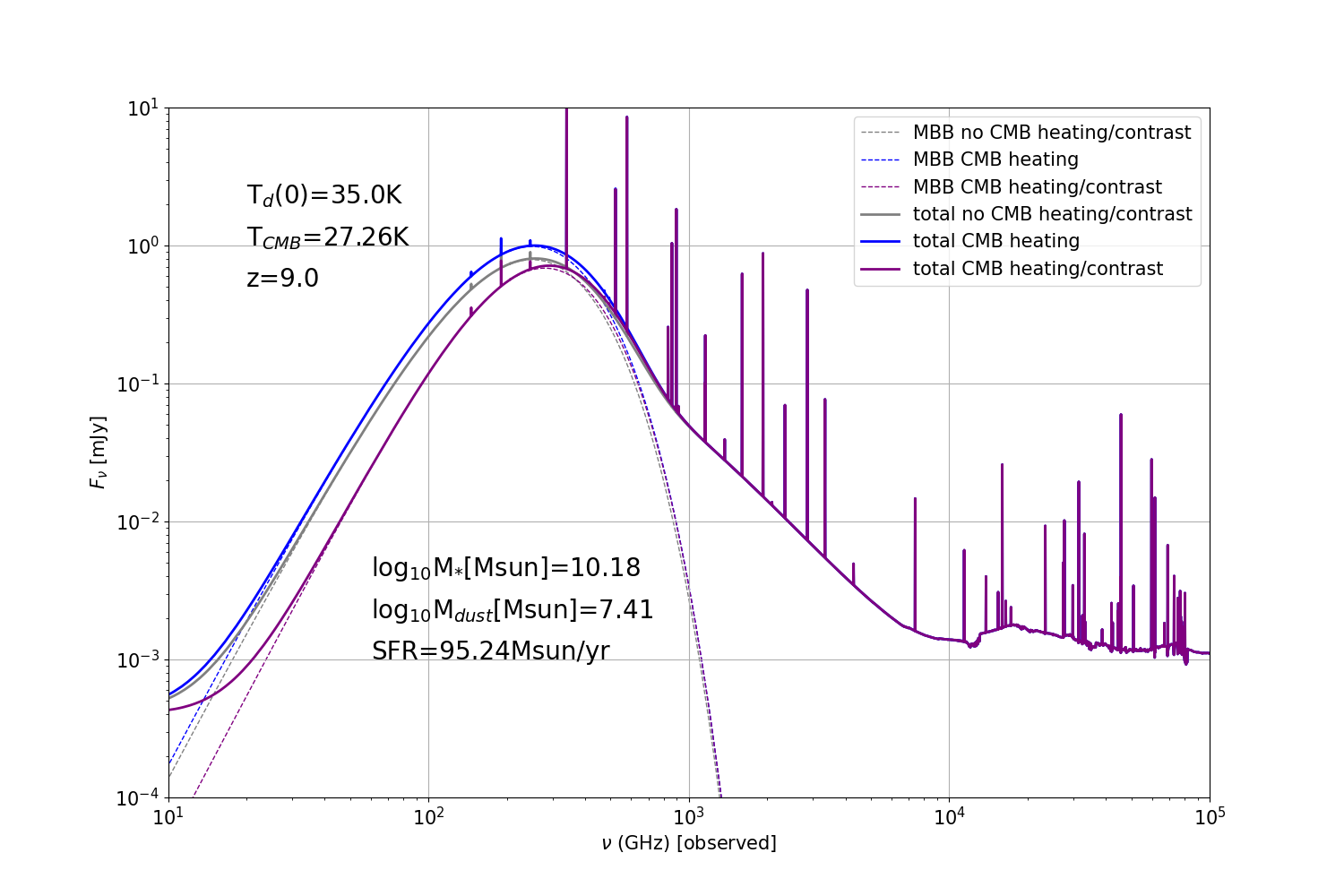}
\caption{Illustration of the creation of model SED with the inclusion of CMB effect. The SED of a model galaxy with $M_{*}=1.5\times10^{10}M_{\odot}$ and SFR=95$M_{\odot}$/yr is created by \texttt{CIGALE} using conventional `delayed' SFH and 36.4K dust temperature at $z=9$ (gray solid line) with 35K dust temperture at $z=0$. Then the additional heating from CMB with $T_{\mbox{\tiny CMB}}=27.26$K at $z=9$ is applied to the modified blackbody function of the SED (blue dashed line), and the modified blackbody function with $T_{\mbox{\tiny CMB}}$ is subtracted to take the background contrast into account (purple dashed line). The `net' SED shown by purple solid line, after taking the CMB effect into account, is the model of an `observed' galaxy SED at high redshift Universe.} 
\label{fig:sed_example}
\end{figure}

Figure~\ref{fig:sed_example} shows the SEDs of a galaxy with $M_{*}=1.5\times10^{10} M_{\odot}$ and dust temperature $T_d(0)=35$K for three different cases: (1) the gray line shows a SED with `elevated' dust temperature $T_d(z)=36.4$K due to the CMB heating at $z=9$ (Equation~\ref{eq:at}) but without having the CMB heating and contrast yet, (2) the blue line shows the same SED including the additional FIR energy from CMB heating with $T_{\mbox{\tiny CMB}}=27.26$K at $z=9$, and (3) the purple line shows the SED after correcting the CMB contrast. The dashed line for each color shows the modified blackbody (MBB) SED component for each case.  

\vspace{22mm}
\bibliography{ALMA_GHZ1}{}
\bibliographystyle{aasjournal}

\end{document}